\documentclass{article}

\usepackage{prime_package}

\usepackage[utf8]{inputenc} 
\usepackage[T1]{fontenc}    
\usepackage{hyperref}       
\usepackage{url}            
\usepackage{booktabs}       
\usepackage{amsfonts}       
\usepackage{nicefrac}       
\usepackage{microtype}      
\usepackage{lipsum}
\usepackage{fancyhdr}       
\usepackage{graphicx}       
\graphicspath{{media/}}     
\usepackage{amsmath}
\usepackage{caption}

\title{Replication of Reference-Dependent Preferences and the Risk-Return Trade-Off in the Chinese Market
}

\author{
  Penggan Xu \\ \\
  School of Management and Economics, \\
  The Chinese University of Hong Kong, Shenzhen \\
  \texttt{pengganxu@link.cuhk.edu.cn} \\
}

\begin{document}
\maketitle

\begin{abstract}
This study replicates the findings of Wang et al. (2017) \cite{wang2017reference} on reference-dependent preferences and their impact on the risk-return trade-off in the Chinese stock market, a unique context characterized by high retail investor participation, speculative trading behavior, and regulatory complexities. Capital Gains Overhang (CGO), a proxy for unrealized gains or losses, is employed to explore how behavioral biases shape cross-sectional stock returns in an emerging market setting. Utilizing data from 1995 to 2024 and econometric techniques such as Dependent Double Sorting and Fama-MacBeth regressions, this research investigates the interaction between CGO and five risk proxies: Beta, Return Volatility (RETVOL), Idiosyncratic Volatility (IVOL), Firm Age (AGE), and Cash Flow Volatility (CFVOL). Key findings reveal a weaker or absent positive risk-return relationship among high-CGO firms and stronger positive relationships among low-CGO firms, diverging from U.S. market results, and the interaction effects between CGO and risk proxies, significant and positive in the U.S., are predominantly negative in the Chinese market, reflecting structural and behavioral differences, such as speculative trading and diminished reliance on reference points. The results suggest that reference-dependent preferences play a less pronounced role in the Chinese market, emphasizing the need for tailored investment strategies in emerging economies.
\end{abstract}

\keywords{Behavioral Finance \and Reference-Dependent Preferences \and Risk-Return Trade-Off \and Capital Gains Overhang (CGO) \and Chinese Stock Market}

\section{Introduction}
The relationship between risk and return is fundamental to finance, yet deviations from classical theories, such as the CAPM, often arise due to behavioral biases. Behavioral finance introduces psychological concepts like reference-dependent preferences and the disposition effect to explain market anomalies. In particular, Capital Gains Overhang (CGO) serves as a key proxy for investor sentiment and risk-taking behavior, reflecting unrealized gains or losses relative to a reference point. Integrating such behavioral biases into factor models can improve the understanding of market anomalies and enhance the predictive power of asset pricing models. Metrics like capital gains overhang (CGO) can serve as explanatory variables, effectively bridging the gap between behavioral finance theories and actionable investment strategies.

This study aims to replicate the key findings of Wang et al. (2017) \cite{wang2017reference} in the context of the Chinese stock market, a rapidly growing and relatively immature market with unique characteristics. The Chinese stock market is distinct due to its high retail investor participation, frequent regulatory interventions, and greater market volatility than developed markets. These unique features provide an ideal environment to test behavioral finance theories, such as reference-dependent preferences, which may manifest differently in a market driven by individual investors. 

The primary objective is to examine whether reference-dependent preferences explain cross-sectional variations in the risk-return trade-off in this emerging market by replicating Wang et al. (2017) \cite{wang2017reference} and comparing the results of the two studies to seek similarities and differences in different markets.

Key empirical findings studied from the Chinese Market include:
\begin{itemize}
    \item Evidence suggests a weaker or absent positive risk-return relationship among high-CGO firms and a stronger positive relationship among low-CGO firms, diverging from U.S. market findings.
    \item Significant but negative interaction effects between CGO and risk proxies (significant and positive in the U.S. market).
\end{itemize}
These findings highlight the reduced role of reference-dependent preferences and the need for tailored investment strategies when conducting factor investing in emerging markets.

This paper is structured as follows: Section \ref{sec:Literature Review} reviews the relevant literature. Section \ref{sec:Methodology} details the methodology, including the definitions of key metrics like CGO \& 5 Risk Proxies, and econometric models (Dependent Double Sort \& Fama-MacBeth Regressions). Section \ref{sec:Empirical Results} presents empirical results, including the CGO Time Series Graph, Dependent Double Sort, and Fama-MacBeth Regressions test results. Section \ref{sec:Discussions and Conclusions} discusses the explanations of the key findings, along with their implications in the context of the Chinese Market. Finally, Section \ref{sec:Future Work} concludes with recommendations for future research.

\section{Literature Review}
\label{sec:Literature Review}
Behavioral finance challenges the traditional view of investor rationality by examining how psychological biases drive market anomalies. Reference-dependent preferences, a key concept in prospect theory introduced by Kahneman and Tversky\cite{kahneman2013prospect}, explain how individuals assess gains and losses relative to a reference point. Wang et al. (2017) \cite{wang2017reference} highlight that investor risk attitudes can vary based on whether their investments are experiencing gains or losses relative to a reference point. This helps to explain deviations from classical models, such as the CAPM, that assume a uniform positive risk-return trade-off. Kőszegi and Rabin (2006) \cite{kHoszegi2006model} further contribute by incorporating dynamic reference points, showing that investors' behavior adjusts as reference points shift.

Frazzini (2006) \cite{frazzini2006disposition} extended this understanding by quantifying the disposition effect—the tendency of investors to sell winning stocks while holding onto losing ones—and introducing the Capital Gains Overhang (CGO) metric. CGO captures the aggregate unrealized gains or losses across investors, offering an empirical proxy for behavioral biases that impact stock pricing. His work highlights that stocks with unrealized gains face selling pressure, while those with unrealized losses remain overvalued. An et al. (2020) \cite{an2020lottery} further examined lottery-like preferences, showing that investors in the loss domain overvalue stocks with high skewness, creating market anomalies. Conversely, investors in the gain domain exhibit risk aversion, diminishing their preference for such stocks. These findings underline the role of reference-dependent preferences in shaping state-dependent risk attitudes.

Despite robust evidence in U.S. markets, limited research on this topic exists for emerging economies like China. The Chinese stock market’s high retail participation, regulatory complexities, and cultural nuances present a unique context to test the universality of these behavioral principles in a different market. This research aims to bridge this gap by replicating key findings from Wang et al. (2017) \cite{wang2017reference}, particularly focusing on reference-dependent preferences and their impact on the risk-return trade-off in the Chinese stock market.

\section{Methodology}
\label{sec:Methodology}
This study adopts a systematic approach to replicate the findings of Wang et al. (2017) \cite{wang2017reference} in the context of the Chinese stock market. Following the original paper, the methodology is designed to examine the role of reference-dependent preferences in shaping the risk-return trade-off using econometric models, behavioral finance metrics, and rigorous data preprocessing techniques.

\subsection{Data Collection and Preprocessing}
\label{subsec:Preprocessing}
The datasets are from several sources, the ones fetched from CSMAR and others provided by the instructor. They comprise stock returns, accounting data, and market-related variables for firms listed on the Chinese stock A-share market, spanning from 1995 to 2024, to ensure robust results. Stocks with less than 10 years of data are excluded from the research to ensure enough data to calculate key metrics that all need at least 2 years of data required by the original paper (in total, there are 1928 stocks left for study). For each time period, stocks that are blacklisted, untradable, and price less than 5 RMB are excluded for calculating the portfolio returns to simulate the actual transaction. Missing values encountered in processes were addressed using interpolation methods for continuous variables and forward-filling for categorical variables, ensuring a complete and reliable dataset. Features were normalized to standardize the scales of different variables, enhancing comparability and preventing any variable from dominating the model due to magnitude differences.

\subsection{Defining Key Metrics}
\label{subsec:Metrics}
\paragraph{CGO}
To measure \textit{\textbf{Capital Gains Overhang (CGO)}}, we first follow what the original paper did, using the turnover-based measure from Grinblatt and Han (2005) \cite{grinblatt2005prospect} to calculate the reference price. At each week t, the reference price for each stock is defined as: 

\begin{equation}
RP_t = \frac{1}{k} \sum_{n=1}^{T} \left( V_{t-n} \prod_{\tau=1}^{n-1} (1 - V_{t-n+\tau}) \right) P_{t-n},
\end{equation}
where $P_t$ is the stock price at the end of week $t$; $V_t$ is week $t$'s turnover in the stock; $T$ is 260, the number of weeks in the previous 5 years; and $k$ is a constant that makes the weights on past prices sum to one. Weekly turnover is calculated as weekly trading volume divided by the number of shares outstanding. The CGO at week $t$ is defined as:

\begin{equation}
CGO_t = \frac{P_{t-1} - RP_t}{P_{t-1}}.
\end{equation}
Finally, to obtain CGO at a monthly frequency, we use the last-week CGO within each month.

\paragraph{Risk Proxies}
To measure risk, five risk proxies were employed following the definition of the original paper, including two main risk proxies (\textbf{\textit{Beta, RETVOL}}) and three alternative risk proxies (\textbf{\textit{IVOL, AGE, CFVOL}}):

\begin{itemize}
    \item \textbf{\textit{Beta:}} using a rolling 5-year regression of the CAPM model. 
    \item \textbf{\textit{Return Volatility (RETVOL):}} the standard deviation of the previous 5-year monthly returns.
    \item \textbf{\textit{Idiosyncratic Volatility (IVOL):}} calculated as the residual volatility from a Fama-French three-factor model. We estimated each stock each month in the data set using the daily return from the previous month.
    \item \textbf{\textit{Firm Age (AGE):}} as a proxy for informational uncertainty, is defined as the time since the firm’s initial listing in the Chinese market.
    \item \textbf{\textit{Cash Flow Volatility (CFVOL):}} the standard deviation of cash flow over the previous 5 years.
\end{itemize}

\subsection{Econometric Models}
\label{subsec:Econometric Models}
Following Wang et al. (2017) \cite{wang2017reference}, the analysis was conducted using two key econometric techniques:

\paragraph{Dependent Double Sorting}
To uncover the heterogeneity in the risk-return relationship based on the levels of CGO and various risk proxies. At the beginning of each month, we divide all firms in our sample into five groups based on lagged CGO, and within each of the CGO groups, we further divide firms into five portfolios based on various lagged risk proxies. This results in a total of 25 portfolios. Each portfolio is then held for one month, and value-weighted excess returns are calculated. Implementing Dependent Double Sorting facilitates identifying patterns and inconsistencies that might not be observable in single-sort analyses. It enables the isolation of interactions between CGO and risk measures, providing more substantial evidence for the hypothesis of reference-dependent preferences. The results are reported in Section \ref{subsec:Dependent Double Sort} in Table 1-5 for five different proxies.

\paragraph{Fama-MacBeth Regressions} 
Although the double-sorting approach is simple and intuitive, it cannot explicitly control for other variables that could influence returns. Since CGO is correlated with other stock characteristics, in particular, past returns and share turnover, the concern could arise that the results in the double sorting are driven by effects other than the capital gains or losses that investors face. To address this critical concern, we perform a series of Fama-MacBeth cross-sectional regressions, allowing us to control for additional variables conveniently. We estimate monthly Fama-MacBeth cross-sectional regressions of stock returns on lagged variables in the following form:
\begin{align}
    R = \alpha &+ \beta_1 \times CGO + \beta_2 \times PROXY \nonumber \\
               &+ \beta_3 \times PROXY \times CGO 
                + \beta_4 \times PROXY \times MOM(-12, -1) \nonumber \\
               &+ \beta_5 \times MOM(-1, 0) 
                + \beta_6 \times MOM(-12, -1) \nonumber \\
               &+ \beta_7 \times TURNOVER + \epsilon
\end{align}
where $R$ is monthly stock return in month $t + 1$, $CGO$ is defined as before at the end of month $t$, $PROXY$ is one of our five risk proxies at the end of month $t$, $MOM(-1, 0)$ is the stock return in month $t$, $MOM(-12, -1)$ is the stock return from the end of month $t - 12$ to the end of month $t - 1$, and $TURNOVER$ is stock turnover in month $t$. The results are reported in Section \ref{subsec:Fama} in Table 6-10 for five different risk proxies.

\section{Empirical Results}
\label{sec:Empirical Results}

\subsection{CGO Time Series Graph}
\label{subsec:CGO Time Series Graph}
Following Wang et al. (2017) \cite{wang2017reference}, we plot the time series of the 10th, 50th, and 90th
percentiles of the cross-section of the CGO generated using Chinese Market data of all individual stocks, reported in below Figure \ref{fig:4_1}:


\begin{figure}[h]
    \centering
    \includegraphics[width=1\linewidth]{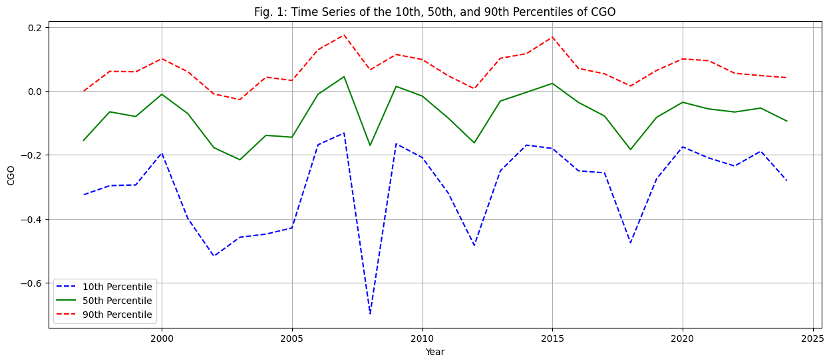}
    \caption{Time series of the 10th, 50th, and 90th percentiles of CGO.}
    \label{fig:4_1}
\end{figure}

Consistent with Figure 1 in Wang et al. (2017) \cite{wang2017reference}, there exists a considerable degree of temporal fluctuation in CGO. Moreover, a significant variation is observed across different entities in CGO, which is crucial for examining the heterogeneity in the risk-return trade-off among firms with varying CGO levels.

\subsection{Dependent Double Sort}
\label{subsec:Dependent Double Sort}
Now, we dive into the key empirical findings of this paper.

Tables \ref{tab:table1}-\ref{tab:table5} below present the empirical results from dependent double sort for the five different risk proxies:

\begin{table}[htbp]
    \centering
    \begin{minipage}{0.48\textwidth}
        \centering
        \small{
        \caption{Portfolio Results with Proxy $= \beta$}
        \label{tab:table1}
        \begin{tabular}{l c c c c}
            \toprule
            Portfolio & CGO1 & CGO3 & CGO5 & Diff-in-Diff \\
            \midrule
            P1 & 1.393 & 1.536 & 1.758 & \\
            P3 & 1.583 & 1.797 & 1.712 & \\
            P5 & 2.306 & 1.649 & 1.785 & \\
            P5 - P1 & 0.913 & 0.113 & 0.027 & -0.886 \\
            $t$-stat & (1.92) & (0.26) & (0.05) & (-2.02) \\
            FF3-$\alpha$ & 0.527 & -0.415 & -0.310 & -0.837 \\
            $t$-stat & (1.52) & (-1.28) & (-0.72) & (-1.84) \\
            \bottomrule
        \end{tabular}
        }
    \end{minipage}
    \hfill
    \begin{minipage}{0.48\textwidth}
        \centering
        \small{
        \caption{Results with Proxy $= \text{RETVOL}$}
        \label{tab:table2}
        \begin{tabular}{l c c c c}
            \toprule
            Portfolio & CGO1 & CGO3 & CGO5 & Diff-in-Diff \\
            \midrule
            P1 & 1.239 & 1.301 & 1.372 & \\
            P3 & 1.859 & 1.520 & 1.715 & \\
            P5 & 2.357 & 2.116 & 2.508 & \\
            P5 - P1 & 1.118 & 0.815 & 1.136 & 0.018 \\
            $t$-stat & (2.32) & (1.84) & (2.05) & (0.033) \\
            FF3-$\alpha$ & 0.753 & 0.265 & 0.890 & 0.137 \\
            $t$-stat & (1.94) & (0.74) & (1.98) & (0.24) \\
            \bottomrule
        \end{tabular}
        }
    \end{minipage}
\end{table}

\begin{table}[htbp]
    \centering
    \begin{minipage}{0.48\textwidth}
        \centering
        \small{
        \caption{Portfolio Results with Proxy $= \text{IVOL}$}
        \label{tab:table3}
        \begin{tabular}{l c c c c}
            \toprule
            Portfolio & CGO1 & CGO3 & CGO5 & Diff-in-Diff \\
            \midrule
            P1 & 1.243 & 1.436 & 1.974 & \\
            P3 & 1.902 & 1.616 & 1.819 & \\
            P5 & 1.601 & 1.476 & 1.493 & \\
            P5 - P1 & 0.358 & 0.041 & -0.481 & -0.839 \\
            $t$-stat & (1.10) & (0.12) & (-1.06) & (-1.80) \\
            FF3-$\alpha$ & 0.453 & -0.103 & -0.788 & 1.241 \\
            $t$-stat & (1.51) & (-0.35) & (-1.79) & (-2.60) \\
            \bottomrule
        \end{tabular}
        }
    \end{minipage}
    \hfill
    \begin{minipage}{0.48\textwidth}
        \centering
        \small{
        \caption{Portfolio Results with Proxy $= 1/\text{AGE}$}
        \label{tab:table4}
        \begin{tabular}{l c c c c}
            \toprule
            Portfolio & CGO1 & CGO3 & CGO5 & Diff-in-Diff \\
            \midrule
            P1 & 1.968 & 1.692 & 1.314 & \\
            P3 & 1.568 & 1.312 & 1.793 & \\
            P5 & 1.591 & 2.066 & 2.248 & \\
            P5 - P1 & -0.377 & 0.374 & -0.934 & 1.311 \\
            $t$-stat & (-1.34) & (1.28) & (2.89) & (3.38) \\
            FF3-$\alpha$ & 0.315 & 0.401 & 0.984 & 1.299 \\
            $t$-stat & (-1.13) & (1.38) & (3.07) & (3.22) \\
            \bottomrule
        \end{tabular}
        }
    \end{minipage}
\end{table}

\begin{table}[htbp]
    \centering
    \small{
    \caption{Portfolio Results with Proxy $= \text{CFVOL}$}
    \label{tab:table5}
    \begin{tabular}{l c c c c}
        \toprule
        Portfolio & CGO1 & CGO3 & CGO5 & Diff-in-Diff \\
        \midrule
        P1 & 2.525 & 2.544 & 1.891 & \\
        P3 & 1.615 & 1.666 & 2.013 & \\
        P5 & 1.325 & 1.330 & 1.509 & \\
        P5 - P1 & -1.200 & -1.215 & -0.382 & 0.818 \\
        $t$-stat & (-3.19) & (-3.11) & (-0.85) & (1.87) \\
        FF3-$\alpha$ & -0.764 & -0.410 & 0.115 & 0.879 \\
        $t$-stat & (-2.42) & (-1.30) & (0.32) & (1.97) \\
        \bottomrule
    \end{tabular}
    }
\end{table}

The results for different risk proxies show mixed results; here, we analyze each table by interpreting it and comparing it with the results and conclusions drawn from the double sort in Wang et al. (2017) \cite{wang2017reference}:

\paragraph{Beta (Table \ref{tab:table1})}
Among high-CGO firms ($CGO5$), the difference between high-beta ($P5$) and low-beta ($P1$) portfolios is $\textbf{\textit{only 0.027\%}}$, and the t-statistic is insignificant ($0.05$). For low-CGO firms ($CGO1$), high-beta firms earn higher returns than low-beta firms ($P5 - P1 = \textbf{\textit{0.913\%}}, t = 1.92$). However, the difference diminishes for medium-CGO and high-CGO firms.

In the original paper, the result for Beta showed a \textbf{\textit{negative risk-return relationship}} among low-CGO firms and a \textbf{\textit{positive relationship}} among high-CGO firms \cite{wang2017reference}. However, in the Chinese market, the expected positive risk-return trade-off for high-CGO firms is \textbf{\textit{largely absent}}, while low-CGO firms show a stronger, positive relationship. This indicates that the Chinese market exhibits weaker or no heterogeneity in the risk-return relationship across CGO levels compared to the original study. This could reflect differences in how systematic risk (Beta) is priced in an emerging market dominated by retail investors.

\paragraph{RETVOL (Table \ref{tab:table2})}
The high-minus-low return difference ($P5 - P1$) among high-CGO firms is \textbf{\textit{1.136\%}}, with a marginally significant t-stat (2.05). However, among low-CGO firms, the risk-return relationship is strongly positive (\textbf{\textit{1.118\%}}, $t = 2.32$), suggesting that high-risk firms outperform low-risk firms even in the loss domain.

In the original paper, \textbf{\textit{negative risk-return trade-offs}} were observed for low-CGO firms, and \textbf{\textit{positive trade-offs}} were found for high-CGO firms \cite{wang2017reference}. In the replication, both low-CGO and high-CGO firms show \textbf{\textit{positive risk-return relationships}}, though the magnitude is more substantial for low-CGO firms. Unlike the U.S. market, the Chinese market appears to \textbf{\textit{reward higher return volatility across CGO levels}}, which may suggest a preference for high-risk, high-return stocks among Chinese investors, irrespective of reference-dependent preferences.

\paragraph{IVOL (Table \ref{tab:table3})}
The difference between high-IVOL and low-IVOL portfolios ($P5 - P1$) is \textbf{\textit{-0.481\%}}among high-CGO firms, with a t-statistic of -1.06, indicating a weak and insignificant negative relationship. Similarly, for low-CGO firms, the relationship is positive but insignificant ($P5 - P1 =$ \textbf{\textit{0.358\%}}, $t = 1.10$).

In the original study, low-CGO firms exhibited a substantial \textbf{\textit{negative risk-return trade-off}}, and high-CGO firms showed a \textbf{\textit{positive relationship}} \cite{wang2017reference}. In the replication, the results are generally \textbf{\textit{weaker}}, with an inconsistent risk-return pattern for both low- and high-CGO firms. The Chinese market does not display clear, robust patterns in the idiosyncratic risk-return trade-off, possibly reflecting market inefficiencies or limited arbitrage due to the dominance of retail investors.

\paragraph{1/AGE (Table \ref{tab:table4})}
Among high-CGO firms, older firms (low 1/AGE) significantly outperform younger firms (high 1/AGE), with a $P5 - P1$ difference of \textbf{\textit{-0.934\%}} ($t = -2.89$). In contrast, for low-CGO firms, younger firms show higher returns than older firms ($P5 - P1$ = \textbf{\textit{-0.377\%}}, $t = -1.34$), though this result is weaker.

The original paper found a positive risk-return relationship among high-CGO firms (favoring older firms) and a negative relationship among low-CGO firms (favoring younger firms) \cite{wang2017reference}. The replication generally aligns with these findings, but the results for low-CGO firms are weaker in the Chinese market. This similarity suggests that \textbf{\textit{firm age}} as a proxy for information uncertainty may play a consistent role across markets, with younger firms being more speculative and risky, particularly in the loss domain.

\paragraph{CFVOL (Table \ref{tab:table5})}
For high-CGO firms, high-CFVOL firms earn \textbf{\textit{-0.382\%}} more than low-CFVOL firms, with a weak t-statistic of -0.85. However, for low-CGO firms, the risk-return trade-off is negative ($P5 - P1 =$ \textbf{\textit{-1.200\%}}, $t = -3.19$), indicating that investors penalize high-risk firms in the loss domain.

In the original paper, low-CGO firms also exhibited a \textbf{\textit{negative risk-return trade-off}}, and high-CGO firms showed a \textbf{\textit{positive relationship}} \cite{wang2017reference}. The replication results for low-CGO firms are consistent with the original findings but with a slightly weaker pattern, as investors strongly penalize high-CFVOL firms in the loss domain. It might suggest that the Chinese market exhibits weaker heterogeneity in the risk-return trade-off for cash flow volatility, likely influenced by its unique investor base and structural characteristics. This contrasts with the more distinct patterns observed in the U.S. market.

\subsection{Fama-MacBeth Regressions}
\label{subsec:Fama}
As argued above, although the double-sorting method is straightforward, it cannot explicitly account for other factors that may impact returns. Given that CGO is associated with various stock attributes, especially past returns and share turnover, the outcomes from the double sorting could be influenced by elements other than the capital gains or losses encountered by investors. To tackle this significant issue, we carry out multiple Fama-MacBeth cross-sectional regressions, enabling us to manage effectively and include additional variables in our analysis. Tables \ref{tab:table6}-\ref{tab:table10} below present the empirical results from Fama-MacBeth regressions for the five different risk proxies (Beta, RETVOL, IVOL, 1/AGE, CFVOL):\\

\noindent {\fontsize{10}{10}\selectfont
\textit{\textbf{Table 6-10:} Fama-MacBeth regressions for 5 different risk proxies.}

\textit{Each month, we run a cross-sectional regression of returns on lagged variables. These tables report the time-series average of the regression coefficients. Variables are defined in Section \ref{subsec:Metrics} and Section \ref{subsec:Econometric Models}. The coefficients are reported in percentages. The sample period is from January 1995 to August 2024. Independent variables are winsorized at 1\% and 99\%. The t-statistics are calculated based on Newey-West adjusted standard errors and reported in parentheses. We use Chinese A-share stocks with a price of at least 5 RMB and are not blacklisted or untradable. The intercept of the regression is not reported.}}\\\\

\begin{table}[htbp]
    \centering
    \begin{minipage}{0.48\textwidth}
        \centering
        \tiny{
        \caption{Fama-MacBeth Regression Results for PROXY $= \beta$}
        \label{tab:table6}
        \begin{tabular}{l c c c c}
            \toprule
            Variable & (1) & (2) & (3) & (4) \\
            \midrule
            CGO & 0.0023 & -0.0004 & 0.1300 & 0.1072 \\
                 & (0.89) & (-0.15) & (6.99) & (5.52) \\
            PROXY & & -0.0103 & -0.0181 & -0.0157 \\
                  & & (-4.08) & (-5.92) & (-4.65) \\
            PROXY $\times$ CGO & & & -0.1172 & -0.0961 \\
                               & & & (-7.34) & (-5.72) \\
            PROXY $\times$ MOM(-12,-1) & & & & -0.0115 \\
                                       & & & & (-1.65) \\
            MOM(-1,0) & 0.0064 & 0.0052 & 0.0059 & 0.0056 \\
                      & (2.03) & (1.66) & (1.82) & (1.69) \\
            MOM(-12,-1) & -0.0178 & -0.0220 & -0.0233 & -0.0140 \\
                        & (-19.67) & (-21.94) & (-22.34) & (-1.67) \\
            TURNOVER & -0.0149 & -0.0153 & -0.0148 & -0.0145 \\
                     & (-3.57) & (-3.92) & (-3.92) & (-3.74) \\
            \bottomrule
        \end{tabular}
        }
    \end{minipage}
    \hfill
    \begin{minipage}{0.48\textwidth}
        \centering
        \tiny{
        \caption{Fama-MacBeth Regression Results for PROXY $= \text{RETVOL}$}
        \label{tab:table7}
        \begin{tabular}{l c c c c}
            \toprule
            Variable & (1) & (2) & (3) & (4) \\
            \midrule
            CGO & 0.0023 & -0.0063 & 0.2694 & 0.2069 \\
                 & (0.89) & (-2.15) & (11.96) & (8.98) \\
            PROXY & & -0.3819 & -0.5270 & -0.4960 \\
                  & & (-13.33) & (-15.62) & (-14.39) \\
            PROXY $\times$ CGO & & & -1.9942 & -1.5470 \\
                               & & & (-12.65) & (-9.80) \\
            PROXY $\times$ MOM(-12,-1) & & & & -0.3854 \\
                                       & & & & (-7.16) \\
            MOM(-1,0) & 0.0064 & 0.0043 & 0.0048 & 0.0031 \\
                      & (2.03) & (1.35) & (1.51) & (0.95) \\
            MOM(-12,-1) & -0.0178 & -0.0214 & -0.0223 & 0.0289 \\
                        & (-19.67) & (-21.30) & (-21.97) & (3.68) \\
            TURNOVER & -0.0149 & -0.0124 & -0.0123 & -0.0132 \\
                     & (-3.57) & (-3.02) & (-3.02) & (-3.12) \\
            \bottomrule
        \end{tabular}
        }
    \end{minipage}
\end{table}

\begin{table}[htbp]
    \centering
    \begin{minipage}{0.48\textwidth}
        \centering
        \tiny{
        \caption{Fama-MacBeth Regression Results for PROXY $= \text{IVOL}$}
        \label{tab:table8}
        \begin{tabular}{l c c c c}
            \toprule
            Variable & (1) & (2) & (3) & (4) \\
            \midrule
            CGO & 0.0023 & 0.0014 & 0.0477 & -0.0121 \\
                 & (0.89) & (0.52) & (9.40) & (-2.25) \\
            PROXY & & 0.1546 & -0.0354 & 0.3459 \\
                  & & (3.85) & (-0.76) & (6.96) \\
            PROXY $\times$ CGO & & & -2.8276 & 0.0508 \\
                               & & & (-9.70) & (0.17) \\
            PROXY $\times$ MOM(-12,-1) & & & & -1.9865 \\
                                       & & & & (-16.71) \\
            MOM(-1,0) & 0.0064 & 0.0096 & 0.0115 & 0.0072 \\
                      & (2.03) & (2.99) & (3.43) & (2.19) \\
            MOM(-12,-1) & -0.0178 & -0.0187 & -0.0182 & 0.0271 \\
                        & (-19.67) & (-20.04) & (-19.04) & (11.15) \\
            TURNOVER & -0.0149 & -0.0169 & -0.0163 & -0.0149 \\
                     & (-3.57) & (-4.01) & (-3.80) & (-3.65) \\
            \bottomrule
        \end{tabular}
        }
    \end{minipage}
    \hfill
    \begin{minipage}{0.48\textwidth}
        \centering
        \tiny{
        \caption{Fama-MacBeth Regression Results for PROXY $= 1/\text{AGE}$}
        \label{tab:table9}
        \begin{tabular}{l c c c c}
            \toprule
            Variable & (1) & (2) & (3) & (4) \\
            \midrule
            CGO & 0.0023 & 0.0041 & -0.0670 & -0.0600 \\
                 & (0.89) & (1.48) & (-6.55) & (-5.62) \\
            PROXY & & 0.1272 & 0.1779 & 0.1572 \\
                  & & (7.19) & (8.63) & (7.35) \\
            PROXY $\times$ CGO & & & 0.5931 & 0.5384 \\
                               & & & (6.71) & (5.60) \\
            PROXY $\times$ MOM(-12,-1) & & & & 0.0513 \\
                                       & & & & (1.44) \\
            MOM(-1,0) & 0.0064 & 0.0058 & 0.0046 & 0.0054 \\
                      & (2.03) & (1.83) & (1.50) & (1.66) \\
            MOM(-12,-1) & -0.0178 & -0.0241 & -0.0254 & -0.0368 \\
                        & (-19.67) & (-21.99) & (-23.22) & (-10.41) \\
            TURNOVER & -0.0149 & -0.0172 & -0.0163 & -0.0160 \\
                     & (-3.57) & (-4.11) & (-4.27) & (-4.16) \\
            \bottomrule
        \end{tabular}
        }
    \end{minipage}
\end{table}

\begin{table}[htbp]
    \centering
    \tiny{
    \caption{Fama-MacBeth Regression Results for Proxy $= \text{CFVOL}$}
    \label{tab:table10}
    \begin{tabular}{l c c c c}
        \toprule
        Variable & (1) & (2) & (3) & (4) \\
        \midrule
        CGO & 0.0023 & 0.0022 & 0.0561 & 0.0366 \\
             & (0.89) & (0.80) & (6.50) & (4.35) \\
        PROXY & & -0.1417 & -0.2395 & -0.2010 \\
              & & (-3.17) & (-4.62) & (-3.64) \\
        PROXY $\times$ CGO & & & -0.1579 & -0.1186 \\
                           & & & (-6.58) & (-4.52) \\
        PROXY $\times$ MOM(-12,-1) & & & & -0.0584 \\
                                   & & & & (-3.14) \\
        MOM(-1,0) & 0.0064 & 0.0056 & 0.0053 & 0.0057 \\
                  & (2.03) & (1.72) & (1.62) & (1.70) \\
        MOM(-12,-1) & -0.0178 & -0.0242 & -0.0255 & -0.0122 \\
                    & (-19.67) & (-23.90) & (-25.15) & (-3.82) \\
        TURNOVER & -0.0149 & -0.0139 & -0.0138 & -0.0125 \\
                 & (-3.57) & (-5.50) & (-5.15) & (-4.69) \\
        \bottomrule
    \end{tabular}
    }
\end{table}

\noindent The results of the Fama-MacBeth regressions show significant differences compared with the original paper across different combinations of variables:

\paragraph{CGO Alone}
In the original paper, the author pointed out that CGO is a key determinant in explaining returns ($\textbf{\textit{1.184, t=7.48}}$): “The benchmark regression in $Column (1)$ shows that the coeﬃcient on CGO is significant and positive, confirming the Fama-MacBeth regression results of Grinblatt and Han (2005)” \cite{wang2017reference}. However, in this research replicating in the Chinese Market, for $Column (1)$, ($\textbf{\textit{0.0023, t=0.89}}$), the result is not as high as the original paper and is insignificant. 

This might suggest that the Chinese market may rely less on reference-dependent preferences tied to capital gains overhang. This could be attributed to structural or behavioral differences between the markets, such as greater speculative behavior or weaker anchoring to past prices among Chinese investors.

\paragraph{Adding Interaction Terms ($\textbf{PROXY*CGO, PROXY*MOM(-12, -1)}$)}
Compared with the original paper, “The results confirm the previous double-sorting analysis that the interaction term is \textbf{\textit{always significant and positive}} for all risk measures, even after controlling for size, book-to-market, past returns, and share turnover,” \cite{wang2017reference} we can see the results in this research for the interaction term are \textbf{\textit{primarily significant and negative}}, except for the $PROXY=1/AGE$.

For $PROXY*CGO$, these findings represent a sharp contrast to the original study. The negative coefficients suggest that $CGO$ may amplify negative risk-return relationships in the Chinese market rather than reinforcing positive ones. This could reflect differences in risk perception, where high-risk stocks are penalized rather than rewarded when coupled with high $CGO$.

For $PROXY*MOM(-12, -1)$, this might suggest that past returns have a more substantial and more persistent moderating effect on the risk-return trade-off in the Chinese market, potentially reflecting a market driven by short-term speculative behavior and higher momentum effects.

\subsection{Limitations}
\label{subsec:Limitations}
\paragraph{Market-Specific Characteristics}
The Chinese stock market is dominated by retail investors, whose behavior differs significantly from the institutional-dominated U.S. market studied in the original paper. This makes direct comparisons challenging and may explain the weaker role of CGO in the Chinese market. Besides, regulatory and structural differences, such as trading restrictions, less stringent financial reporting requirements, and the presence of government-controlled entities, may also influence the results and limit the generalizability of findings to other emerging markets.

\paragraph{Data and Sample Period}
The study's results are based on data from the Chinese market over a specific sample period (1995-2024), which may not capture longer-term trends or structural changes in the market as in the U.S. (the original paper used 1964-2014 \cite{wang2017reference}).

\paragraph{Measurement of CGO}
While the methodology for calculating CGO was replicated from Wang et al. (2017), it may not fully capture reference points or investor expectations in the Chinese market. For instance, cultural and psychological factors may lead Chinese investors to use different reference points, such as psychological price levels or speculative benchmarks. Besides, the use of turnover-based measures may be affected by the high trading volume and speculative nature of the Chinese market, potentially biasing CGO calculations.

\section{Discussions and Conclusions}
\label{sec:Discussions and Conclusions}
Based on the replication of the Double Sort, this research underscores significant differences between the Chinese and U.S. markets. The weaker or absent heterogeneity in the risk-return trade-off across CGO levels suggests that reference-dependent preferences may be diminished in the Chinese context. Structural market features, such as the dominance of retail investors, speculative behavior, and market inefficiencies, likely contribute to these deviations from the original findings. While certain proxies, such as firm age and CFVOL, show similarities across markets, others, like Beta and IVOL, reveal contrasting dynamics. 

Based on the replication of the Fama-MacBeth regressions, the results suggest that the heterogeneity in the risk-return trade-off observed in the U.S. is much less pronounced in the Chinese market. Structural and behavioral differences, such as the dominance of retail investors, speculative trading practices, and limited anchoring to reference points, likely contribute to the observed deviations. 

The findings of this research underscore the necessity of tailoring factor-based investment strategies to specific market contexts. While behavioral finance theories provide a robust framework for understanding investor behavior, their application must consider the unique characteristics of individual markets. In the case of the Chinese market, the diminished role of CGO, the more potent momentum effects, and the contrasting risk-return dynamics point to the need for customized approaches that align with local investor preferences and market structures.

\section{Future Work}
\label{sec:Future Work}
While this study highlights significant differences in the risk-return trade-off between the Chinese and U.S. markets, the above limitations underscore the need for further refinement of the models and methods used. By addressing these issues through robustness checks and additional analyses, future research can provide more definitive insights into how market structure, investor behavior, and cultural factors influence the heterogeneity of risk-return dynamics in the Chinese market. 

To address the identified limitations and enhance the robustness of the findings, the following steps are recommended for future research:

\paragraph{Alternative CGO Measures}
Explore different methodologies for calculating CGO, such as mutual fund holdings or metrics adjusted for speculative trading behavior, to better reflect reference points used by Chinese investors.

\paragraph{Incorporation of Extended Control Variables}
Introduce additional variables like investor sentiment indices, leverage ratios, and liquidity measures to address potential omitted variable bias and refine the analysis of CGO's effects.

\paragraph{Exploration of Alternative Models}
Utilize advanced statistical and machine learning techniques to capture nonlinearities and interaction effects more effectively, providing deeper insights into the dynamics of CGO and risk proxies.

\paragraph{Institutional vs. Retail Investor Impact}
Conduct separate analyses of stocks dominated by institutional versus retail investors to investigate how investor composition influences CGO's role in the risk-return trade-off.

\bibliographystyle{unsrt}  
\bibliography{references}

\end{document}